# On the Influence of the Laser Illumination on the Logic Cells Current Consumption

## *First measurement results*

Dmytro Petryk[1], Zoya Dyka[1,2], Milos Krstic[1,3], Jan Bělohoubek[4,5], Petr Fišer[4], František Steiner[5], Tomáš Blecha[5], Peter Langendörfer[1,2] and Ievgen Kabin[1]

[1] *IHP – Leibniz-Institut für innovative Mikroelektronik,* Frankfurt (Oder), Germany
[2] *BTU Cottbus-Senftenberg,* Cottbus, Germany
[3] *University Potsdam,* Potsdam, Germany
[4] *Czech Technical University in Prague,* Prague, Czech Republic
[5] *University of West Bohemia,* Pilsen, Czech Republic

{petryk, dyka, kabin, krstic, langendoerfer}@ihp-microelectronics.com, {petr.fiser, jan.belohoubek}@fit.cvut.cz, {belohoub, steiner}@fel.zcu.cz

*Abstract*—**Physical side-channel attacks represent a great challenge for today's chip design. Although attacks on CMOS dynamic power represent a class of state-of-the-art attacks, many other effects potentially affect the security of CMOS chips analogously by affecting mostly static behaviour of the chip, including aging, ionizing radiation, or non-ionizing illumination of the CMOS. Vulnerabilities exploiting data dependency in CMOS static power were already demonstrated in practice and the analogous vulnerability exploiting light-modulated static power was demonstrated by simulation. This work confirms the CMOS vulnerability related to the light-modulated data-dependent static power experimentally and discusses future work.**

I. INTRODUCTION

Evaluating the impact of laser illumination, ionizing radiation, and accelerated aging on transistors applied in various electronic systems is an important task from a reliability and security point of view as well as to ensure the compliance with industry standards. Electronic systems that are exposed to laser radiation or irradiation can experience Single Event Effects (SEE), increase in the noise floor [1], or increase of the power consumption, which can cause a transient or permanent change in the system's behaviour leading to potential reliability or security issues.

The sensitivity to SEEs can increase with ionizing radiation dose or accelerated aging. This can lead to system failures or malfunctions unacceptable in critical applications such as aerospace, defence, or medical electronics. The ionizing radiation causes permanent changes in silicon, forming defects called interface traps. Interface traps commonly affect the transistor threshold voltage, and increase the flicker noise floor [1], and could potentially impact the reliability of cryptographic primitives. In crypto-primitives based e.g. on static random access memories (SRAMs), degradation defects may cause entropy loss of a true random number generator (TRNG) or degradation of physically unclonable functions (PUFs), or even lead to disclosing novel attack vectors related to side-channel analysis in CMOS in general.

The sensitivity of semiconductor devices to environmental parameters such as temperature, light, etc., can be exploited to perform physical attacks. For example, laser illumination can be used as a means for fault injection attacks on electronic systems if the systems are physically accessible. By analysing the changes in the system's behaviour due to laser radiation, an attacker can potentially extract sensitive information such as encryption keys or other confidential data. It has already been shown that illumination could decrease the side-channel analysis complexity by simulation [2] and by measurement [3]. The results of the experiment reported in [3] do not provide evidence of significant improvement in the attack complexity (measured as the number of traces leading to the successful attack), but it was designed as a highly ad-hoc and black-box experiment. According to our previous theoretical results [2] when such an attack would (i) be complemented with knowledge about the chip's layout, and (ii) target the circuit static power, the impact would be significant [2]. By evaluating the impact of laser illumination on transistors, designers and engineers can better understand the system's susceptibility to these attacks. Having accurate models has a paramount importance, as they enable attack mitigation by supporting efficient countermeasures to be realized in the design process.

Many industries have established standards for the evaluation of electronic systems' resistance to laser radiation, such as the aerospace standard MIL-STD-883 or the telecommunications standard ITU-T K.45. By evaluating the impact of laser on transistors, designers and engineers can ensure that their electronic systems comply with these standards.

The reaction of semiconductor devices on laser illumination can significantly be influenced by aging processes. Thus, aging can – possibly – not only significantly improve side-channel analysis attacks but also enable new fault injection techniques for attacks against semiconductor devices. The influence of aging on the transistor behaviour, including the transistor reaction to the laser illumination, has to be investigated and included in the theoretical transistor behaviour models.

Aging can be caused by exposure to various environmental conditions like temperature, moisture, irradiation, and electrical stress. Accelerated aging is then used to predict reliability. This is no different for semiconductors. Many conventional procedures are used to evaluate the aging of semiconductors, such as Time-Dependent Dielectric Breakdown (TDDB), Hot Carrier Injection (HCI), Bias Temperature Instability (BTI), or Electro-migration (EM) [4]. In the latest research [5] a high-efficiency aging test technique is presented.

The theoretical models can provide valuable insights into the behaviour of the transistor under aging or laser illumination independently, but they can also be used to predict the transistor response under the combination of different laser parameters, total ionizing radiation dose, ionization intensity, aging, etc. These models are typically based on a combination of experimental data and theoretical calculations and require a deep understanding of the physical and electrical properties of the materials and devices involved. The practical evaluation of the theoretical models requires the development of experimental setups that can accurately measure the transistor response, for example under laser illumination.

In this paper, we present the first experimental proof of static power data dependency of logic cell under laser illumination, describe the limitations connected with the measurement setup, and show the fundamental correctness of the preliminary simulation results experimentally [2]. This paper is structured as follows. Section II recaps how the laser beam could be exploited for side-channel attacks. Section III describes the chip under test. Section IV overviews the setup used in the experiments. Section V gives results of the experiments. Section VI discusses open questions and future work. Section VII concludes the paper.

II. LASER-INDUCED STATIC POWER DATA DEPENDENCY

The recent work has lit up the data dependency of the optical beam induced current (OBIC), induced by a focused laser beam [2]. According to [2], OBIC amplifies the data-dependent static power of the laser-beam-covered CMOS area by a factor of 4-5 compared to the data-dependent leakage [6]. This fact could be potentially exploited to compromise a CMOS circuit by applying a (more) efficient combined side-channel attack to the laser-beam-covered CMOS area. To understand the nature of the potential vulnerability, models based on the work related to Photoelectric Laser Stimulation (PLS) [7] were created. These models allow SPICE simulation of the induced OBIC in the simple bulk CMOS structures under PLS. CMOS logic is composed of complementary NMOS/PMOS stacks, where, simply said, parallel structures are complemented by serially arranged structures and vice versa. This causes different nodal voltages to appear in the circuit nodes under different input configurations causing different OBICs to be induced under different circuit input configurations.

**Fig. 1** shows the basic mechanics of the OBIC in a CMOS structure comprising a stack of three MOS transistors.

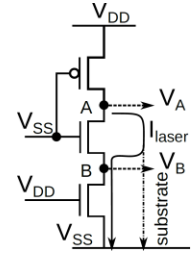

**Fig. 1.** Basic mechanics of the OBIC in the CMOS structure under laser illumination.

In the transistor stack shown above, $V_B \approx 0$, as the lower transistor is ON; $V_A \approx V_{DD}$, as the upper transistor is ON and the middle transistor is OFF. In this configuration, the n+/p-sub junction in node A is reverse-biased. When the CMOS stack is illuminated, OBIC is generated at the PN junction in node A. The current loop is closed through node B and the substrate.

We performed extensive electrical simulations for a set of basic logic cells and observed a strong dependency of the induced OBIC on input cell pattern, and also on the illumination power, as shown in **Fig. 2** for the two-input NAND cell[1] [2]. Such an induced data-dependent OBIC from the illuminated cells increases the power consumption of the whole chip. The relation between the illuminated area and the area of the whole chip defines significantly the "visibility" of OBIC contribution in the power trace of the whole chip. If the illuminated area is small, the OBIC contribution will be insignificant or even negligible. We expect that such data-dependent OBIC contributions would be hidden in the noise for larger chips, but still, it could potentially be a source of information leakage when statistical processing is employed. Our models[2] could be tailored for a SPICE simulation of custom bulk CMOS logic. The disadvantage of our models is that they were validated on simple bulk CMOS structures and they do not take into account complex layout-induced effects like reflections.

---

[1] The output (Out) of a NAND gate will only be false (a logic '0') if input 1 (In1) and input 2 (In2) are true (a logic '1'), i.e. Out = $\overline{In1 \cdot In2}$. Further, in the paper, we denote two input pattern as two binary digits, e.g. '00', where the first digit represent logic state at In1 and the second – at In2.

[2] https://github.com/DDD-FIT-CTU/CMOS-PLS

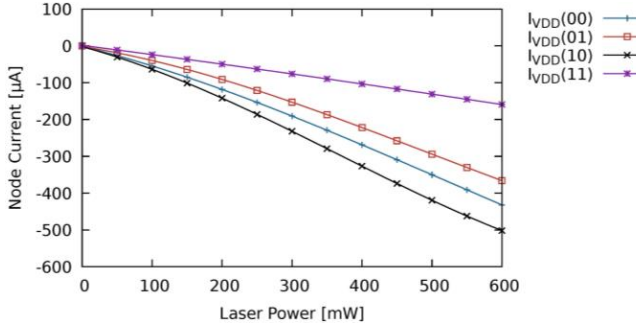

**Fig. 2.** The power imprint of the NAND2X1 cell under different laser powers and input patterns (in brackets in the legend) in TSMC180.

## III. CHIP UNDER TEST

We have run our first experiments using chips manufactured in the IHP's 250 nm technology [9]. Originally, the chips were designed to measure signal propagation delays through different types of cells. We denote these chips as Libval025. They were manufactured in the "old" technology with only 5 metal layers that allowed to manufacture chips without metal fillers[3]. Each Libval025 chip has wires only in the two bottom metal layers and no metal fillers over its cells. Thus, any area of the chip can be freely illuminated with a laser.

Libval025 has a pulse generator, and contains the following cell types: NAND, NOR, Inverter, and Flip-flops. The cells are connected in chains and each contains 4 lines with cells of a single type, e.g., a sequence of NAND cells, or a sequence of inverters with different loads, i.e., the chip has 16 lines in total.

Only one out of the 16 chains was active in the Libval chip during the experiment; other lines were deactivated by selection pins. A more detailed description of the Libval025 chip can be found in [10]. The structural scheme of Libval025 is shown in **Fig. 3**.

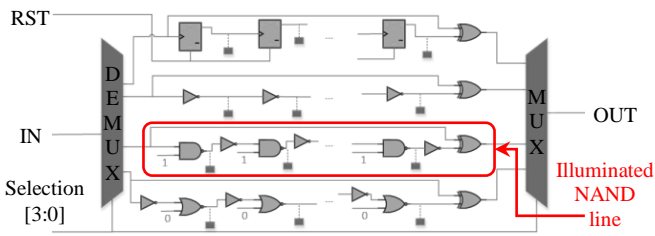

**Fig. 3.** Structural scheme of Libval025.

We experimented with a line of NAND cells in the Libval chip. Please note that one of the inputs of each NAND cell is constant '1'. Hence, when illuminating the chip, we were able to experiment only with two different input combinations of the NAND cell, i.e., '01' and '11'.

## IV. EXPERIMENTAL SETUP

### A. Equipment

To evaluate the response of transistor(s) to the laser illumination we have used the equipment available at IHP. The setup is schematically shown in **Fig. 4**.

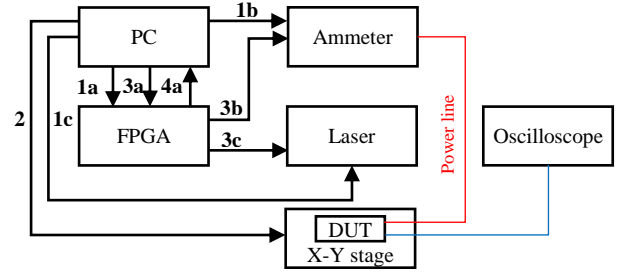

**Fig. 4.** Experimental laser illumination setup.

It consists of (i) a PC used for equipment configuration and experiment control; (ii) a precise Keysight [11] ammeter for measurements of current in the range of femto Amperes; (iii) a red single-mode laser by Alphanov [12]; (iv) an FPGA [13] for precise laser and ammeter synchronization; (v) a high-precision X-Y stage [14]; (vi) an oscilloscope [15] used to monitor the circuit output; and (vii) a 1st generation Riscure Diode Laser Station [16] and a microscope with a 50× magnification objective, used to shrink the laser beam spot size, connected to the laser via optical fibre.

We started each experiment with an initialization step sending configuration parameters from the PC to the FPGA (step 1a), the ammeter (step 1b), and to the laser (step 1c). For the ammeter, the current measurement range, sampling rate and trigger parameters were pre-configured. For the laser, its beam pulse power, the pulse duration as well as the trigger for the pulse start/end were also pre-configured. The parameters of the laser can be found in [8]. The Libval025 chip is the device under test (DUT) placed on the X-Y stage. The moving speed of the X-Y stage is controlled and set using the PC (step 2). The selection of the NAND cell area for the laser illumination was done manually. In step 3a the PC sends a signal to the FPGA to start the experiment. The FGPA receives the start signal from the PC and sends a signal to the ammeter to start the current measurements (step 3b). After that the FGPA sends a signal to the laser to perform the laser illumination (step 3b). The signal is sent with a small time delay, i.e., the laser is activated during ongoing measurements. In step 4a the FPGA sends the signal "ready" to the PC after the current measurement is finished, whereby the time of the current measurements is set to be longer than the laser pulse duration. Measured data is saved using the ammeter.

### B. Setup configuration and initial parameters

The evaluation of the cell response to the laser illumination is a complex experimental task, due to the small measured currents. To ensure accurate measurement results, even with state-of-the-art equipment, all possible external noise sources have to be eliminated. To keep the noise level low during measurements, we powered the Libval chip with standard AA batteries via a voltage divider. The voltage divider is used to limit the current drawn from power supply for reducing the likelihood of chip damage and to ensure different voltages needed to power the chip. According to our preliminary evaluations, changes of current drawn from a power supply equal to dozens of pA can be observable using our setup.

---

[3] Small metal structures that are placed in different metal layers to comply with technology manufacturing process requirements. The metal fillers are obstacles for laser illumination and can reduce the success rate of laser fault injection attacks [8].

To ensure the control of the laser source and the ammeter, a custom software was written. The control commands were implemented according to [17] and [18]. To reduce the stress on the chip, we configured the equipment based on the minimal time between measurements of the ammeter that amounts to 10 µs [11]. According to our observations, the minimal time between measurements is not stable and varies from 11 µs to 12 µs[4]. To avoid excessive heating of the device and reduce the time of increased current flow through the cell under illumination in our experiments, we:

- used a laser beam pulse duration of 100 µs and increased the power with a minimal step of 1 % from the minimal power[5] of 1 % till we observed a laser influence;
- performed measurements within a 300 µs time interval without laser illumination. To measure the current, we set 25 sample points in the trigger settings of the ammeter that gives us 25·11 µs = 275 µs ≤ measurement time interval ≤ 25·12 µs = 300 µs.

To target only several transistors of the NAND cell, we used a 50× magnification objective. **Fig. 5** shows the area of NAND cells and the laser spot size. The laser beam spot diameter using a 50× magnification objective is around 2.6 µm[6]. The spot is measured in continuous mode using the laser beam profiler from Kokyo [19], see **Fig. 6**.

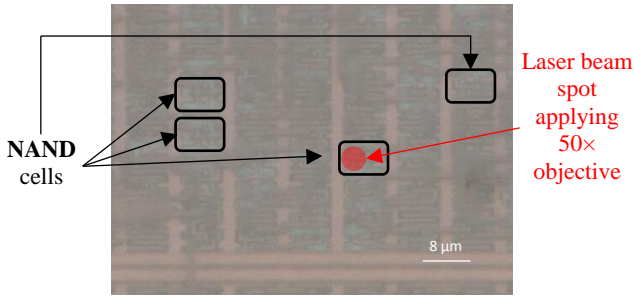

**Fig. 5.** The illuminated area (marked red) of the NAND cell in the Libval025 chip (applying a 50× magnification objective).

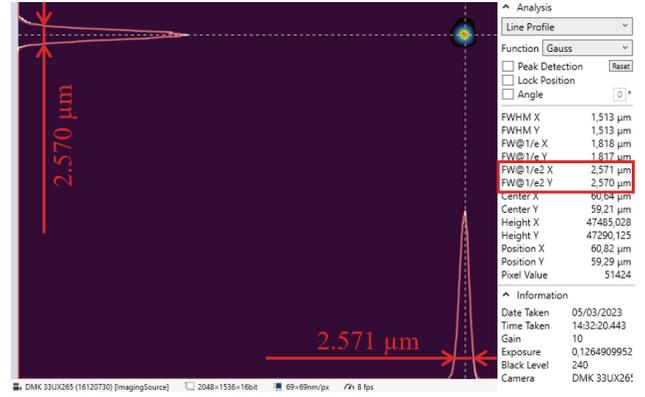

**Fig. 6.** Laser beam spot size of the single-mode laser using a 50× magnification objective.

During measurements, the sample has been placed into a black-box to protect the user from possible harmful laser reflections, and to supress unwanted external illumination. We illuminated the NMOS transistor area primarily, the X-Y stage has been switched off to eliminate movements of the sample.

V. EXPERIMENTAL RESULTS

Using our laser illumination setup, we illuminated a small area of the NAND logic cell[7]. We were able to measure the current drawn from the power supply by the attacked circuit under the cell illumination and without the laser illumination.

In more detail, we measured the current flowing through the core power supply line, i.e., the line that powers the logic cells. Since one of the inputs of each two-input NAND cell in the Libval025 chip is always set to logic '1', we measured the current drawn from the power supply with only two input state combinations: '0', '1' and '1', '1'. According to the measurements, the influence of the laser is clearly observable with 7 % laser beam power without injecting any fault, i.e., without affecting the output of the combinational logic.

Without laser illumination, the measured operating current drawn from the power supply is in the range of about ±0.1 µA, see the blue line in **Fig. 7**. Illuminating the area of NMOS transistors using 7 % laser beam power we observed an increase of the current up to 0.75 µA with logic input '11' (see green line in **Fig. 7**), and to 0.81 µA with logic input '01' (see red line in **Fig. 7**). We repeated the experiments 3 times and obtained similar results.

---

[4] It was only achievable by setting the following minimal values: 0.005 power clock cycles and 10 µs trigger period.
[5] To ensure smoother increase of laser beam output power we used only 1 from 2 channels available of the used laser.
[6] It is the diameter of the laser beam spot measured at $1/e^2$ (≈13.5 %), i.e. it is a spot in which 86.5 % of laser beam intensity is concentrated.
[7] The size of the NAND cell is 35 µm$^2$ of which about 5 µm$^2$ is illuminated by laser beam.

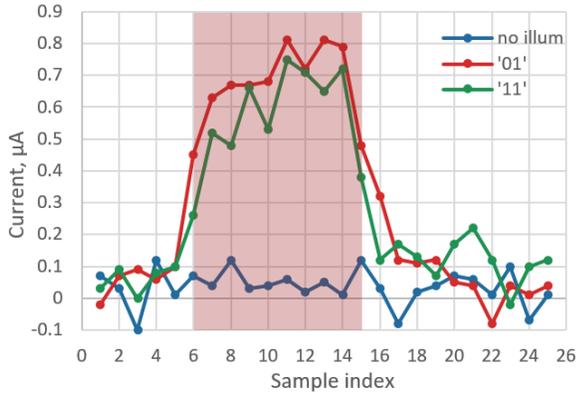

**Fig. 7.** Measured currents while illuminating NAND cells with logic inputs '01' and '11'.

The results of our measurements clearly show that under the laser illumination, the current drawn from the power supply depends on the input values of the illuminated NAND cell, i.e. the current is higher when the cell inputs are '01' than for '11', which is consistent with the simulations presented in [2].

Since the first significant increase of the current happened at sample index 6, we interpret it as the start point of the laser influence, see **Fig. 7**. According to the time stamps obtained from the ammeter, the time between *sample index 6* and *sample index 15* equals to 99 μs (for the cell inputs '11') and 100 μs (for the cell inputs '01'). This time interval is marked red in **Fig. 7**.

Additionally, we performed experiments with increased laser beam output powers: 10 %, 15 % and 20 %. We measured a significant increase of the current applying the increased laser beam power, see **Fig. 8**. In the experiments with the increased laser beam power, we did not observe a change of the chip's output state, i.e., no faults were injected.

The measured current when applying 15 % and 20 % laser beam output power is quite similar. We assume that a kind of "saturation" is reached. We did not measure the current illuminating the NAND cell with input '11' with 20 % laser beam output power to avoid potentially damaging the chip.

The realized experimental setup allows to practically evaluate the influence of OBIC on the side-channel leakage, which was investigated theoretically in [2].

## VI. OPEN QUESTIONS AND FUTURE WORK

We see the necessity of further measurements and the need for chips specially designed for further measurements, allowing accurate characterization of the theoretical models of the cell reaction on the laser illumination, for a given technology node. Such models can provide the opportunity to simulate attacks based on laser illumination at early stages of cryptographic chip development and, consequently, to develop appropriate security countermeasures. Additionally, this can save the financial costs required for redesigning and manufacturing if a design was successfully attacked. In addition, it can be used as part of a methodology for engineering chips resistant to a wide range of physical attacks. There is a need for accurate models considering complex layout-induced effects allowing accurate simulations, and also for correct simplified models allowing fast simulations of big circuits.

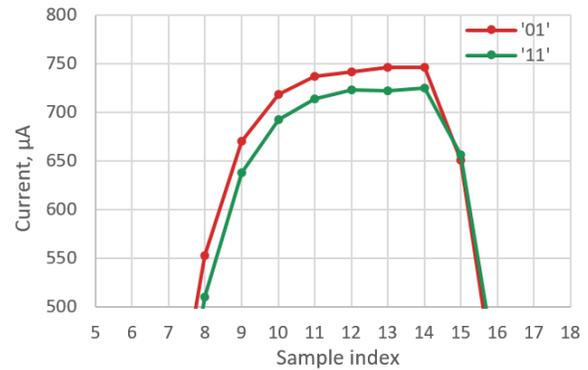

*(a)*

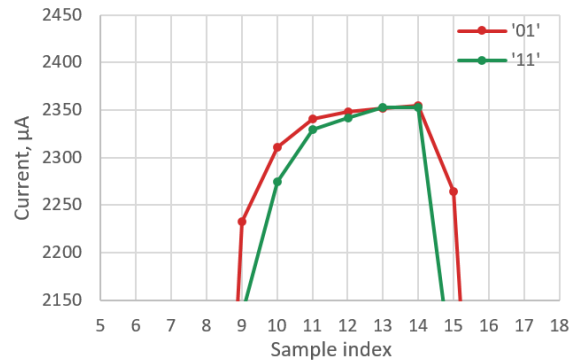

*(b)*

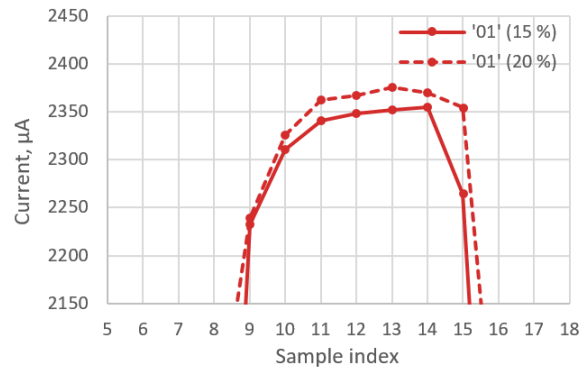

*(c)*

**Fig. 8.** Measured currents illuminating the NAND cell applying the following laser beam power: *(a)* – 10 %, input '01' and '11'; *(b)* – 15 %, '01' and '11'; *(c)* – 15 % and 20 %, input '01'.

The effects related to aging, accelerated aging, or ionizing radiation in CMOS might have analogous effects from the security perspective alone or if they are combined with the CMOS illumination. The impact of those effects on reliability is widely studied, their security consequences, namely the effects on side-channel analysis attacks or effects on parameter degradation of security primitives are uncovered in many cases and need further research. From the security perspective, the significant results cover only the circuit vulnerability with respect to static power side-channel attacks [6], and this paper in particular confirms the simulation results only for illumination alone combined with no other effects.

VII. CONCLUSION

We presented the first successful experiments influencing the static power consumption of NAND cells by laser illumination. In the experiments carried out, it was feasible to increase the current drawn from the power supply significantly. According to the measured data, the magnitude of the current depends on the laser beam power and the logic state of the NAND cell inputs. When applying an increased laser beam power, the measured current increases significantly (up to ≈2.4 mA). The results are consistent with simulations presented in [2] and confirm that optical beam induced current (OBIC) could be potentially exploited to compromise a CMOS circuit.

Our observation shows that a narrow laser beam can still influence other cells due to internal reflections. Hence, to evaluate the influence of laser illumination on a single cell, and specifically to characterize models accurately, special chips have to be designed and manufactured for additional experiments.